\documentclass[12pt]{article}

\usepackage[pdftex]{graphicx}
\usepackage[absolute]{textpos}

\usepackage{amssymb}
\usepackage{amsmath}

 \newcommand{\dpar}[2] { \frac{\partial {#1} } {\partial {#2}} }
 \newcommand{\ndpar}[3] { \frac{\partial^{#3} {#1} } {\partial {#2}^{#3} }}

\newcommand{\schr}{Schr\"o\-din\-ger}
\newcommand{\nlse}{nonlinear \schr\ equation}

\newcommand{\ivp}{initial-value problem}

\newcommand{\bc}{boundary condition}

\newcommand{\bs}{boundaries}
\renewcommand{\P}{periodic}
\newcommand{\eNss}{exact $N$-soliton solution}

\newcommand{\s}{soliton}

\newcommand{\Enteros}{\mathbb{Z}}

\title{Forces on Solitons in Finite, Nonlinear, Planar Waveguides}

\author{J. I. Ramos and F. R. Villatoro \\
Departamento de Lenguajes y Ciencias de la Computaci\'on \\
E. T. S. Ingenieros Industriales \\
Universidad de M\'alaga \\
Plaza El Ejido, s/n \\
29013-M\'alaga  \\
SPAIN \\
}

\date{}

\begin{document}

\maketitle

\begin{abstract}

The forces acting on
and the energies of solitons governed by the \nlse\ in  finite, planar
waveguides with periodic and with homogeneous Dirichlet, Neumann and Robin
boundary conditions are determined by means of a quantum analogy. It is shown
that these densities have $S$-shape profiles and increase as the hardness of the
boundary conditions is increased.
\end{abstract}

\vspace{1cm}

KEY TERMS:   Spatial solitons, \nlse, finite planar waveguides

\section{Introduction}

\label{se:introduction}

Soliton propagation has received a great deal of
attention in  recent years because of its possible applications in
distorsionless signal transmission in ultra-high speed and long-distance
transoceanic telecommunications by optical fibers~\cite{Moll88}, soliton lasers,
all-optical soliton switches~\cite{Tril91}, nonlinear planar
waveguides~\cite{Bian91}, etc. However, most of the theoretical studies on
one-dimensional solitons have been concerned with the initial-value problem,
i.e., infinite intervals in space, or with boundary-generated solitons,
i.e., semi-infinite intervals, and have been based on
the inverse scattering transform~\cite{Ablo81}, perturbation methods or the
infinite sine Fourier transform~\cite{Foka89}.

Since nonlinear transmission lines, optical fibers, etc., have a finite length,
the infinite and semi-infinite spatial
domains used in most previous studies of nonlinear wave propagation are
idealizations. In this letter, the \nlse\ in finite intervals with periodic and
homogeneous Dirichlet, Neumann and Robin boundary conditions is studied
numerically in order to determine the energy of and the forces on solitons
prior to, during and after their interaction with the boundaries. The
energy of and the forces on solitons are evaluated by means of a quantum
analogy between wave propagation phenomena governed by the \nlse\ and the
motion of a particle in a nonlinear potential well. Bian and Chan~\cite{Bian91}
have previously studied the nonlinear and diffraction forces on solitons
 governed
by  the initial-value problem of the \nlse. These authors, however, used a
filter  and their force and energy densities are not consistent with a strict
quantum mechanics analogy.

\section{One-Dimensional, Nonlinear Schr\"{o}dinger 
Equation}

The envelope of the electric field in a two-dimensional medium with a
quadratic nonlinear refractive index, $n$=$n_0+n_2|E'|^2$, i.e., subject to the
Kerr effect, with neither losses nor other higher order effects, can be modelled
by the following one-dimensional \nlse~\cite{Hase73}
\begin{equation} \label{eq:optnse}
2ik\dpar{E'}{z'} + \ndpar{E'}{x'}{2} + k^2 \frac{n_2^2}{n_0^2} |E'|^2 E' = 0
\end{equation}
where $E'$ is the envelope of the electric field, $k$ is the
wave number  in a linear medium with $n$=$n_0$, $x'$ and $z'$ are spatial
 coordinates,
 and $n_2$ represents the nonlinear part of the refractive
index. Within
certain power constraints, the solution of Eq.~\eqref{eq:optnse} is a spatial
soliton which can propagate in a slab waveguide  of width  much larger than the
transverse mode size of the soliton.

Using the following changes of variables
$E'$=${n_0}/{k n_2} u$,
$z'$=$ 2 k t$,
$x'$=$x$,
Eq.~\eqref{eq:optnse} can be written in the following
dimensionless form
\begin{equation} \label{eq:nse}
i u_{t} = -u_{xx} + V u,  \qquad x\in{\cal{D}},  \quad t \geq 0
\end{equation}
where $V$=$ - | u |^2$ and $\cal D$ is the spatial domain.

Equation~\eqref{eq:nse} can be written in the standard, dimensional form
\nlse\ by transforming the indepedent variables as indicated in~\cite{Bian91}.
In this letter, however, the dimensionless Eq.~\eqref{eq:nse} is used to draw an
analogy between the propagation of a soliton in a waveguide and the motion of a
particle in a nonlinear potential wave. This analogy allows to determine the
dimensionless quantum momentum and quantum energy of and the dimensionless forces
on the soliton as follows. The linear quantum momentum is defined by the
one-dimensional operator~\cite{Land77}
\begin{equation} \label{eq:mom:op}
P \equiv - i  \dpar{}{x}
\end{equation}
whose mean value is given by the following time-dependent expression
\begin{equation} \label{eq:mom:mv}
\langle P \rangle = \frac { \langle u,P u \rangle } {\langle
u, u \rangle}
= - i  \frac { \int_{\cal D} u^* u_{x} \,d{x} }
{\int_{\cal D} u^* u\,d{x } }.
\end{equation}
where $u^*$ denotes the complex conjugate of $u$.

The quantum momentum density is defined as
\begin{equation} \label{eq:mom:ld}
p(x,t) = -i u^* u_{x}.
\end{equation}
The linear quantum energy is defined by the operator
\begin{equation} \label{eq:ene:op}
E = i  \dpar{}{t}
\end{equation}
whose mean value is
\begin{equation} \label{eq:ene:mv}
\langle E \rangle = \frac { \langle u,E u \rangle }
{\langle u, u \rangle}
=  i  \frac { \int_{\cal D} u^* u_{t} \,d{x} }
{\int_{\cal D} u^* u\,d{x} }.
\end{equation}
while the local energy density is
\begin{equation} \label{eq:ene:ld}
e(x,t) = i  \frac {u^* u_{t}}{\varrho}.
\end{equation}
where~$\varrho$ is a factor which has been introduced in order to unify our
treatment with that of Bian and Chan~\cite{Bian91} even though these authors
used the dimensional \nlse\ to analyze the forces on solitons propagating in
infinite lines.

A physically and mathematically consistent use of the quantum
analogy can be obtained by employing the following renormalization factor
\begin{equation} 
\varrho = {\int_{\cal D} u^* u\,dx }
\end{equation}
which coincides with the first invariant of the \ivp\ of the \nlse.
Since depending on the applications of the \nlse\ and boundary conditions, the
first invariant may not remain constant, a value $\varrho$=1 is used in this
work.

The energy density can be split  as
\begin{equation} \label{eq:lde:comp}
e(x,t) = -\frac {u^*u_{xx} + |u|^4}{\varrho} = e_k(x,t) + e_v(x,t)
\end{equation}
where the  kinetic and potential energy densities are, respectively,
\[
e_k(x,t) = -\frac {u^*u_{xx}}{\varrho}, \qquad
e_v(x,t) = -\frac {|u|^4}{\varrho}.
\]

From the potential energy density, the following nonlinear
force density that produces  self-focusing on solitons is obtained
\begin{equation} \label{eq:lde:ev2}
f_n(x,t) = - \dpar{}{x} e_v(x,t)
= \dpar{}{x} \left( \frac{|u|^4}{\varrho} \right)
\end{equation}
while, from the kinetic one, the following diffraction force density that is
responsible  for the diffraction effect on the \s\ is obtained
\begin{equation} \label{eq:lde:ev3}
f_d(x,t) = - \dpar{}{x} e_k(x,t)
= \dpar{}{x} \left( \frac {u^*u_{xx}}{\varrho} \right) .
\end{equation}

The energy and force densities defined above have complex
values. In quantum mechanics, these complex values
do not represent any problem because only the mean values of the energy and
force can be  physically measured and these values are always real due to the
Hermitian quantum  operators.
In order to obtain physical insight  from the energy and force densities
defined above, it is  convenient to use their real values.

The force densities may be  employed to determine their effects on solitons.
Bian and Chan~\cite{Bian91} used the filter $\varrho$=$u^*u$
in order to assess  that a soliton can be considered as the result
of two opposite  phenomena, i.e.,  the diffraction and the self-focusing caused
by  dispersion and nonlinearity, respectively, for the \ivp\
of Eq.~\eqref{eq:nse}. Note that the force densities obtained by Bian and Chan
are somewhat artificial and are not consistent with the quantum mechanics analogy
since their renormalization factor is a function of both space and time, and,
therefore, affects the values of these densities in a different manner
depending on the soliton location at each instant of time.

The numerical  results of Bian and Chan~\cite{Bian91} for the \ivp\ of
the \nlse\ indicate that  the total force density, i.e., the sum of the
diffraction and nonlinear force  densities, vanishes for the 1-soliton
solution.  For the general \eNss~\cite{Zakh72,Gord83}, their definition  yields
a non-zero total force which is controlled by the nonlinear one
indicating that \s s are compressed while propagating.

In this letter, finite planar waveguides in a symmetric, finite interval $\cal
D$=$[-L,L]$ subject to the following homogeneous boundary conditions are
considered
 \begin{equation} \label{eq:nse:r}
u(-L,t) + \gamma u_x(-L,t) = 0, \qquad u(L,t) + \gamma u_x(L,t) = 0,
\qquad t \geq 0.
\end{equation}
The values $\gamma$=0 and $\infty$ correspond to homogeneous Dirichlet and
Neumann boundary conditions, respectively. The \nlse\ with homogeneous Robin
boundary conditions is a  skew-symmetric problem in $x$ except for the
limiting cases $\gamma$=0 and
 $\infty$ for which it is
symmetric; therefore, the interaction  of a soliton with the left boundary is
different from that with a right one if $\gamma$ is finite and different from
zero. In this letter, mixed boundary conditions corresponding to $\gamma$=1 are
considered, and the results for the Dirichlet, Neumann and Robin boundary
conditions are compared with those for the initial-value problem and with those
for the following periodic boundary conditions  \begin{equation} \label{eq:nse:p}
\ndpar{u}{x}{n}(x,t) = \ndpar{u}{x}{n}(x+2kL,t),  \qquad \forall n \geq 0, \quad
k \in \Enteros,  \quad x \in {\cal D} \equiv [-L,L], \quad t \geq 0.
\end{equation}

\section{Presentation of Results}

In this section, some sample results (cf. Figs.~1--5) that illustrate the
nonlinear force density, the real part of the diffraction force density, the
total force density, and the real part of the momentum density are presented as
functions of space and time  for the four types of boundary conditions
considered in this letter. Figures~1--5 also show the space-time isocontours of
the three-dimensional data presented in these figures. The results presented in
Figs.~1--5 were obtained by means of a Crank-Nicolson finite-difference method
and correspond to an interval of $L$=50, amplitude and velocity of the soliton
equal to one, initial position and initial phase of the soliton equal to zero,
and spatial and temporal step sizes equal to 0.25 and 0.01, respectively. These
figures show that the nonlinear and diffraction force densities have an
$S$-shape, that  the  nonlinear force density is larger than the diffraction one
when the \s\ is far away from the boundaries,  and that the momentum density has
a  bell-shape similar to that of the soliton amplitude.

The boundary does not affect the force and momentum densities on the \s\ for the
\P\ \bc s as shown in Fig.~1. However, the force
densities are strongly affected by the \bs\ for the Robin boundary
conditions and the momentum changes  sign since the soliton
rebounds from the boundary, recovering the shape that it had
prior to the collision. Figure~2 corresponds to Dirichlet boundary conditions
and shows that the $S$-shape of the force densities is maintained during the
collision process although very large forces are reached in the collision
process  since the \s\ steepens near the boundary  because its amplitude  is
zero there; the momentum at the boundary is exactly zero and changes sign
smoothly.

Figure~3 corresponds to homogeneous Neumann boundary conditions and illustrates
the very large values reached by the force densities  and the change in their
shapes during the  collision; the total force density exhibits extrema and the
diffraction force exceeds the magnitude of the nonlinear one at the boundary.

Since the \nlse\ with Robin boundary conditions is not a symmetric problem, the
interaction of a soliton with the left boundary is expected to be different from
that with the right one. This is clearly illustrated in Figures~4 and~5 which
correspond to the first and second collisions, i.e., a collision with the right
boundary followed by another one with the left boundary.

Figure~4 indicates that the diffraction force is larger than the nonlinear one
only near the boundary resulting in a positive total force. The momentum density
behaves similarly to the Neumann case, except that its value
at the right boundary is small but different from zero. Figure~5 shows that the
nonlinear (diffraction) force density exhibits a plateau near the boundary
during the collision before changing its $S$-shape and reaching a negative
(positive) extremum at the boundary. Figure 5
 indicates that  the nonlinear
force is  always greater than the diffraction one when soliton collides with
the left boundary, indicating that the collision with the left boundary is
similar to that observed with Dirichlet boundary conditions (cf. Fig. 2). The
isocontours presented in Figures 4 and 5 indicate that the soliton becomes
closer to the right boundary than to the left one.


\section{Conclusions}

A quantum mechanics analogy is used to determine the forces on and the energies
of solitons in finite, nonlinear planar waveguides subject to periodic and to
homogeneous Dirichlet, Neumann and Robin boundary conditions in finite
intervals. For all the boundary conditions considered in this letter, it has
been  shown that the nonlinear, diffraction and total force densities have
$S$-shape  profiles and recover the values that they had prior to the
interaction with the
 boundary.  The interaction of the soliton with the boundary is characterized by
force densities which increase as the hardness of the boundary conditions is
increased.


\section*{Acknowledgments}

This research was supported by the Spanish D.G.I.C.Y.T. under Project  no.
PB91--0767. The second author (F.R.V.) has a fellowship from the
Programa Sectorial de Formaci\'on de Profesorado Universitario y Personal
Investigador, Subprograma de Formaci\'on de Investigadores "Promoci\'on General
del Conocimiento", from the Ministerio de Educaci\'on y Ciencia of Spain.



\newcommand{\bookref}[6]{#1, ``{\em {#2}}." #3, #6, p. #5.}

\newcommand{\paperref}[6]{#1, ``#2," {\em {#3}}, Vol. #4, #6, pp. #5.}

\newcommand{\procref}[7]{#1, ``#2" in ``{\em #3}" edited by #4, #5, pp. #6,
#7.}

\newcommand{\procrefvol}[9]{#1: ``#2" in ``{\em #3}", #4, edited by #5, #6, vol.
#7, pp. #8, #9.}



\newpage
\thispagestyle{empty}
\begin{textblock*}{\paperwidth}(0mm,0mm)
   \noindent\includegraphics[width=\paperwidth,height=\paperheight]{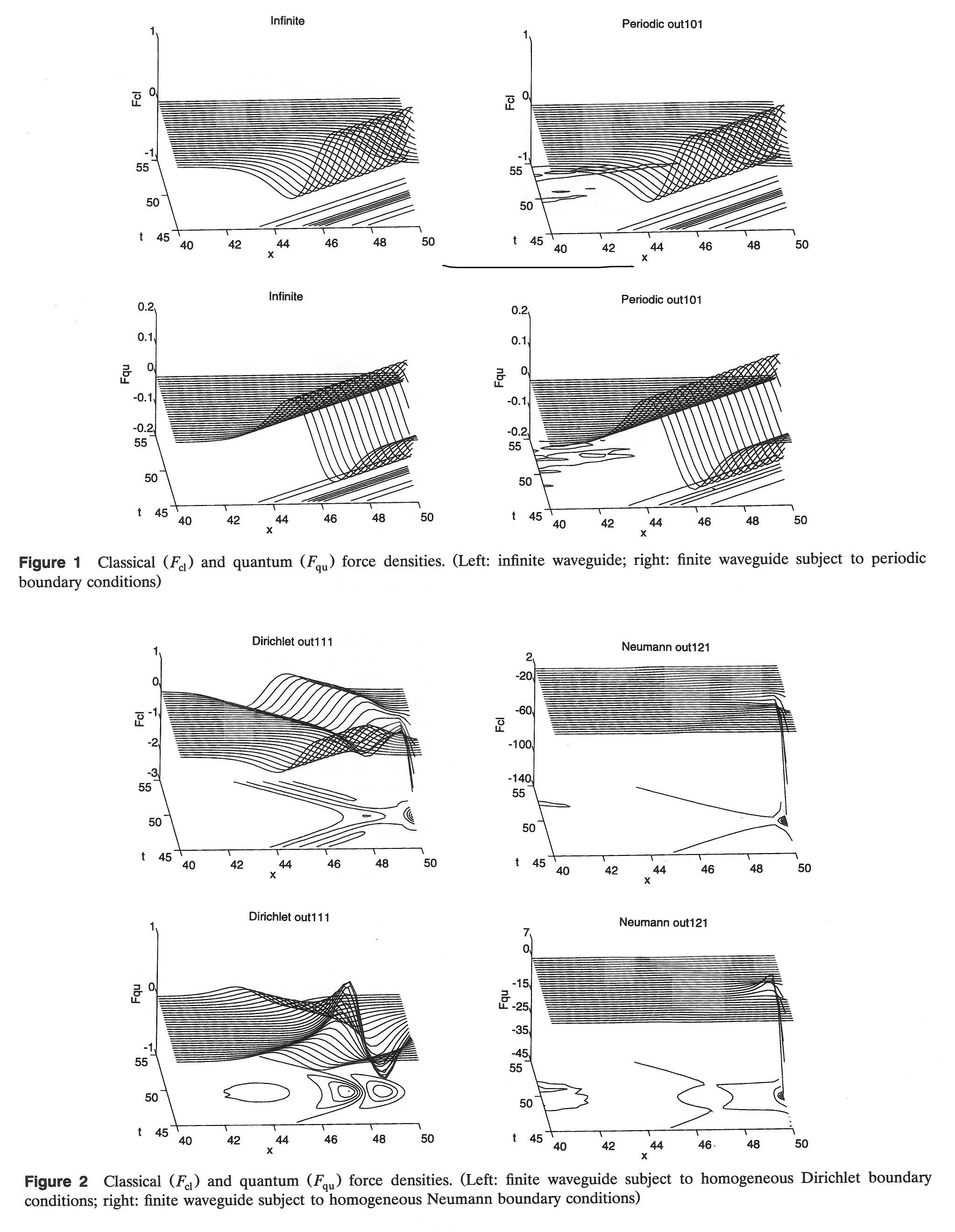}
\end{textblock*}
\mbox{}\newpage

\newpage
\thispagestyle{empty}
\begin{textblock*}{\paperwidth}(0mm,0mm)
   \noindent\includegraphics[width=\paperwidth,height=\paperheight]{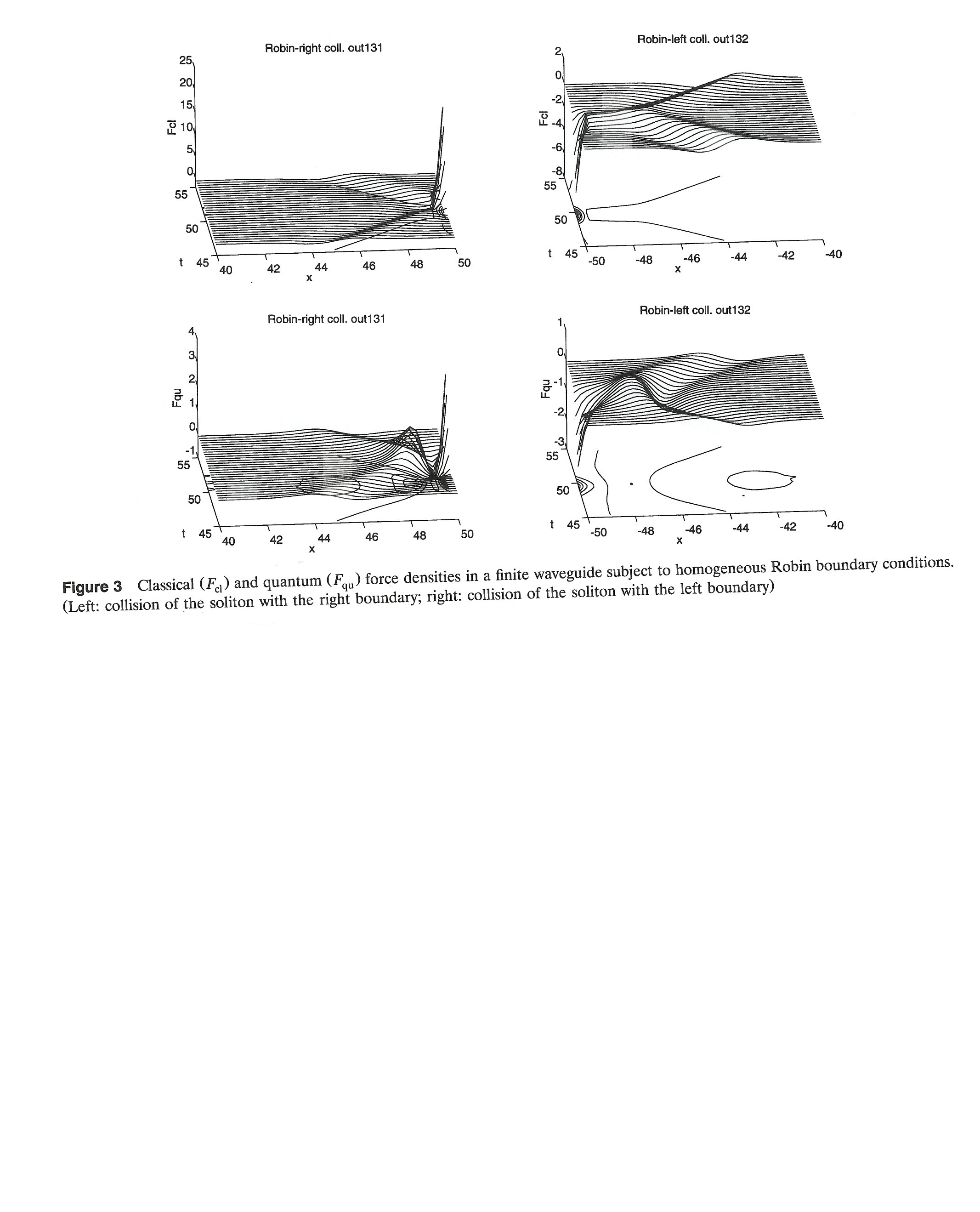}
\end{textblock*}
\mbox{}\newpage

\end{document}